\documentclass[preprintnumbers,showpacs,preprintnumbers, twocolumn,
amsmath,amssymb,APSl,prd,nofootinbib,superscriptaddress]{revtex4-2}

\usepackage{graphicx} % Required for inserting images
\usepackage{xcolor}
\usepackage{amsmath}
 \usepackage{hyperref}
\usepackage[top = 2cm, bottom = 2cm, right = 2cm, left =2cm]{geometry}
\usepackage{subcaption}
\newcommand{\be}{\begin{equation}}
\newcommand{\ee}{\end{equation}}
\newcommand{\bea}{\begin{eqnarray}}
\newcommand{\eea}{\end{eqnarray}}
\newcommand{\beaa}{\begin{eqnarray*}}
\newcommand{\eeaa}{\end{eqnarray*}}

\newcommand{\nn}{\nonumber \\}
\newcommand{\e}{\mathrm{e}}

\allowdisplaybreaks[4] 
\begin{document}

\preprint{KEK-TH-2742, KEK-Cosmo-0387}

\title{Non-stationary wormholes with the presence of scalar fields and modified gravity}

\author{G.~Alencar}
\email{geova@fisica.ufc.br}
\affiliation{Departamento de F\'isica, Universidade Federal do Cear\'a, Caixa Postal 6030, Campus do Pici, 60455-760 Fortaleza, Cear\'a, Brazil}

\author{R.~D\'{a}rlla}
\email{robertadarlla2@gmail.com}
\affiliation{Departamento de F\'isica, Universidade Federal do Cear\'a, Caixa Postal 6030, Campus do Pici, 60455-760 Fortaleza, Cear\'a, Brazil}

\author{Shin'ichi~Nojiri}
\email{nojiri@nagoya-u.jp}
\affiliation{KEK Theory Center, Institute of Particle and Nuclear Studies,
High Energy Accelerator Research Organization (KEK), Oho 1-1, Tsukuba, Ibaraki 305-0801, Japan}
\affiliation{Kobayashi-Maskawa Institute for the Origin of Particles and the Universe, Nagoya University, Nagoya 464-8602, Japan}

\author{Sergei~D.~Odintsov}
\email{odintsov@ice.csic.es} \affiliation{Institut de Ci\`{e}ncies de l'Espai,
ICE/CSIC-IEEC, Campus UAB, Carrer de Can Magrans s/n, 08193 Bellaterra (Barcelona),
Spain}
 \affiliation{Instituci\'o Catalana de Recerca i Estudis Avan\c{c}ats (ICREA),
Passeig Luis Companys, 23, 08010 Barcelona, Spain}

\author{Diego~S\'aez-Chill\'on~G\'omez}
\email{diego.saez@uva.es} 
\affiliation{Department of Theoretical Physics, Atomic and Optics, and Laboratory for Disruptive Interdisciplinary Science (LaDIS), Campus Miguel Delibes, \\ University of Valladolid UVA, Paseo Bel\'en, 7,
47011 - Valladolid, Spain}
\affiliation{Departamento de F\'isica, Universidade Federal do Cear\'a, Caixa Postal 6030, Campus do Pici, 60455-760 Fortaleza, Cear\'a, Brazil}

%\date{\today}

\begin{abstract}
Some novel regular spacetimes are considered that show a non-stationary wormhole structure. 
A Simpson-Visser-like procedure is applied to reconstruct these regular spacetimes, free of time-like and space-like singularities. 
Such a procedure is also applied to describe a regular cosmological expansion, where the universe reaches a minimum scale and then rebounds. 
This type of regular spacetime is achieved by considering some scalar fields as sources, with the appropriate kinetic term and scalar potential. 
We show that all these sources become ghosts due to the wrong sign of the kinetic term. 
Nevertheless, these ghosts can be eliminated by constraints. The same procedure is also explored in the framework of modified gravities, particularly within the so-called 
$f(R)$ gravity, where a new wormhole spacetime is also obtained that does not require a ghost scalar field.
\end{abstract}

\maketitle

%%%%%%%%%%%%%%%%%%%%
\section{Introduction}
%%%%%%%%%%%%%%%%%%%%

Over the last few years, General Relativity (GR) and Cosmology have gained significant attention. 
This is due to the present era of high-precision measurements of black holes and cosmological effects, which allowed the direct measurement of gravitational waves by LIGO, 
VIRGO and KAGRA, the image of supermassive black holes by the Event Horizon 
Telescope (EHT), precise of the cosmological parameters and recent works that investigate detection methods for wormholes in both general relativity and alternative theories of gravity \cite{LIGOScientific:2016aoc, EventHorizonTelescope:2022wkp, Planck:2018vyg, ACT:2020gnv, DeFalco2020}. 
In GR (and other theories of gravity), wormholes are hypothetical solutions that describe regular spacetimes, featuring a tunnel-like structure that connects different regions/universes. 
Most of these types of spacetime known in the literature are static spherically symmetric solutions of 
Einstein's field equations with the presence of additional fields that generally do not satisfy the energy conditions \cite{Visser:1995cc, Lobo:2017book, morris1988wormholes}. 
Wormholes function as shortcuts that connect two different regions of the spacetime or even different spacetimes, since, unlike black holes, wormholes do not have event horizons.

The first wormhole-like solution was the so-called Schwarzschild wormhole, also known as Einstein-Rosen bridge~\cite{RosenEinstein}, which is based on the 
Schwarzschild metric, where the maximal extension of such a spacetime leads to two separate regions/universes that might be connected through a tunnel. 
However, these wormholes are highly unstable. 
Small disturbances in spacetime would cause the structure to collapse instantly, preventing any safe passage \cite{wheeler1955geons, instability_wormholes}. 
Another type of wormhole is the Morris-Thorne wormhole~\cite{Morris1988WormholesIS}, which, unlike the Schwarzschild wormhole, is traversable. 
However, for this wormhole to remain open and stable, it would require the presence of exotic matter, a type of theoretical matter with negative energy density. 
These theoretical objects have been extensively studied in the literature \cite{visser2003traversable,Lobo:2005us,Lobo:2005yv, Li:2020jyf}. 
Recent studies have focused largely on the metrics of black bounces \cite{Simpson_2019, lima2022black, Furtado:2022tnb,Lima:2023arg,Lima:2023jtl,Alencar:2024yvh,Alencar:2025jvl,Rois2025}, which are spacetimes that might transit from regular black holes into traversable wormholes, depending on a free parameter that characterizes the behavior of the spacetime. 
One of the procedures for reconstructing black bounce solutions was proposed by the 
Simpson-Visser~\cite{Simpson_2019} and then implemented in a plethora of different spacetimes~\cite{lima2022black, Furtado:2022tnb, 
Lima:2023arg, Lima:2023jtl, Alencar:2024yvh, Alencar:2025jvl}. 
These interesting objects have gained attention as their properties and dynamics have been explored in depth, also in some extensions of 
GR~\cite{Radhakrishnan:2024rnm, Elizalde:2023rds, DeFalco2023, Garcia:2010xb, Lobo:2008zu,Lobo2009a,Harko2013}. 

One of the most studied exact solutions for these structures is the Ellis-Bronnikov wormhole~\cite{Ellis:1973yv, Bronnikov:1973fh}. 
Unlike black holes, the Ellis-Bronnikov wormhole does not have a central singularity, which makes it an interesting, regular, and theoretical solution for the study 
of hypothetical non-stationary objects. 
Such a spacetime is a specific solution of Einstein's field equations in the presence of an exotic type of matter, often called phantom matter with negative energy density, 
which violates the zero energy condition and enables the stability of the structure. 
This type of exotic matter can be described by a free scalar field with negative kinetic energy. 
Nevertheless, other wormhole solutions can be reconstructed with several scalar fields that might avoid the presence of ghosts and, consequently, no instabilities arise \cite{Nojiri:2023dvf}. 

In this work, a new non-stationary solution is found based on the Ellis-Bronnikov wormhole by exploring the types of sources responsible for this solution. 
To do so, firstly, a time regularization is considered for Friedmann-Lema\^itre-Robertson-Walker (FLRW) spacetime in order to reconstruct cosmologies free of time-like singularities, where a 
Simpson-Visser-like procedure is applied \cite{Chataignier:2022yic}, leading to bouncing cosmological models \cite{Bamba:2013fha, Nojiri:2017ncd, delaCruz-Dombriz:2018nvt}. 
Then, in order to construct a FLRW cosmology free of singularities, a particular scalar field model, called cuscuton, is considered \cite{Afshordi:2006ad}. 
The cuscuton field has been widely analyzed previously in the literature, since it lacks any dynamical degree of freedom, and it might reproduce not only 
the late-time accelerating expansion of the universe \cite{cuscuton}, but also the inflationary phase \cite{Bartolo:2021wpt, Dehghani:2025udv}. 
Here, the presence of the cuscuton field transforms singular cosmologies into regular ones. 
The procedure is applied to construct a non-stationary Ellis-Bronnikov wormhole that is consequently free of any type of singularities, and its throat expands/contracts with time, 
reaching a minimum size of the throat and then bouncing. 
Nevertheless, we show that in order to obtain such regular solutions, the scalar field, including the cuscuton field, has a negative kinetic term, turning into a ghost field 
and consequently, the model might not be physically acceptable. However, this reconstruction procedure might be applied to any other type of spacetime to make it regular. 
Furthermore, we show that the ghosts can be eliminated by constraints. 
Finally, such analysis is extended to $f(R)$ gravities \cite{Nojiri:2017ncd}. 
To do so, the same regular solutions are attempted to be held in $f(R)$ gravities. 
Cosmological bouncing solutions are well known in modified gravities, where a regular cosmological expansion can be easily reconstructed \cite{Bamba:2013fha}. 
Moreover, while some wormhole solutions are known to exist in the framework of $f(R)$ gravity (for a recent review, see \cite{Radhakrishnan:2024rnm}), we have revisited the 
Ellis-Bronnikov wormhole in the framework of $f(R)$ gravity \cite{Lobo2009a} and found that the solution is not satisfied when considering the same scalar field as in 
GR, but other types of sources are necessary. 
In addition, a new wormhole spacetime is obtained in the framework of $f(R)$ gravity, which has similar properties to the 
Ellis-Bronnikov wormhole, but does not require the presence of a ghost field. 

This paper is organized as follows: in Section~\ref{timeregu}, a brief review of FLRW cosmologies is followed, and the cuscuton field is introduced as a source for regularizing the metric. 
Then, Section~\ref{NSEllisBronnikov} is devoted to the reconstruction of a new non-stationary and regular wormhole solution departing from the Ellis-Bronnikov wormhole, where the scalar fields implied in the solution, including the cuscuton, become ghosts. 
In Section~\ref{ghostsection}, we show that the ghosts can be eliminated by constraints. 
In Section~\ref{FRsection}, these regular solutions are analyzed in the framework of $f(R)$ gravity and a new solution is found. 
Finally, Section~\ref{conclusions} summarizes the main results of the paper.

%%%%%%%%%%%%%%%%%%%%%%%%%%
\section{Time Regularization for FLRW cosmologies} \label{timeregu}
%%%%%%%%%%%%%%%%%%%%%%%%%%

The Friedmann-Lema\^itre-Robertson-Walker (FLRW) model describes an isotropic and homogeneous expanding universe that characterizes the geometry of the universe. 
In comoving coordinates, the metric can be written as follows
\begin{align}
ds^2 = -dt^2 + a^2(t)\left[\frac{dr^2}{1-K r^2} + r^2 d\Omega^2 \right]\, ,
\label{frlw}
\end{align}
where $a(t)$ is the scale factor and $d\Omega^2=d\theta^2+\sin^2\theta d\varphi^2$, which expresses the metric of a two-dimensional sphere with a unit radius, whereas $K$ defines the spatial curvature. The dynamics of this model are described by the FLRW equations, which can be obtained directly from Einstein's field equations, leading to 
\begin{align}
H^2= \left(\frac{\dot{a}}{a}\right)^2 =&\, \frac{\kappa^2}{3}\rho - \frac{K}{a^2}+\frac{\Lambda}{3}\, , 
\label{friedmann1}\\
\frac{\Ddot{a}}{a} =&\, \frac{\kappa^2}{6}(\rho+3p) + \frac{\Lambda}{3} \, ,
\label{friedmann2}
\end{align}
where $\kappa^2=8\pi G$, $\rho$ is the energy density, $p$ is the pressure and $\Lambda$ is a cosmological constant. The causal structure of the FLRW model is an important aspect of understanding how different regions of the universe can influence each other over time 
and is strongly influenced by the evolution of the scale factor $a(t)$.  This scale factor governs the expansion of the universe and defines the limits of the cosmological horizons, both the future and the past light cones, 
such that they delineate the regions of the universe that are ``observable'' from a given point in spacetime and consequently regions  that are causally connected in the past and the future. 
Nevertheless, this causal structure generally has a singularity at the beginning of the universe, where one might choose $t=0$. For the standard model of cosmology, solutions to the equations (\ref{friedmann1}) and (\ref{friedmann2}) make the density $\rho$ diverges as $t \rightarrow 0$, leading to the 
Big Bang singularity, where the scale factor $a(t)$ turns out to be zero, and the temperature becomes infinite. Indeed, this is the case when the universe is filled with any kind of matter that satisfies the energy conditions. In order to find a source that regularises the FLRW spacetime, a cuscuton-type scalar field, $\phi_1$, is considered. The action for this scalar field is given by \cite{cuscuton}
\begin{align}
S_{\phi_1} = \int d^4x \sqrt{-g} \left[\mu \sqrt{2X}-U \left(\phi_1\right) \right]\, ,
\label{actionCusco}
\end{align}
where $\mu$ is a constant that might be either positive or negative, whereas,
\begin{align}
\label{X}
X = -\frac{1}{2}g^{\mu\nu}\partial_\mu \phi_1 \partial_\nu \phi_1\, .
\end{align}
Here we assume $X>0$, that is, the vector $\left( \partial_\mu \phi_1 \right)$ is time-like. 
The equation of motion can be obtained by varying the action (\ref{actionCusco}) with respect to the scalar field $\phi$, where one obtains
\begin{align}
\frac{1}{\sqrt{-g}} \partial_{\gamma} \left[ \frac{\sqrt{-g}\partial^{\gamma}\phi_1}{\sqrt{-g^{\alpha \beta}\partial_{\alpha}\phi_1 \partial_{\beta}\phi_1}} \right] 
 - \frac{1}{\mu} \frac{dU}{d\phi_1} = 0\, .
\label{motion}
\end{align}
In addition, the variation of the action $S_{\phi_1}$ over the metric $g_{\mu\nu}$ provides the energy-momentum tensor for the cuscuton field:
\begin{align}
T^\mu{}_\nu =&\, \frac{\mu g^{\mu\gamma}\partial_{\gamma}\phi_1\partial_{\nu}\phi_1}{\sqrt{-g^{\alpha \beta}\partial_{\alpha}\phi_1 \partial_{\beta}\phi_1}} \nonumber \\
&\, + \delta^{\mu}_{\nu}\left[\mu\sqrt{-g^{\gamma \alpha} \partial_{\gamma}\phi_1 \partial_{\alpha}\phi_1} - U\left( \phi_1 \right) \right]\, ,
\label{tensormecuscuton}
\end{align}
whose components are:
\begin{align}
T^t{}_t =&\, - U\left(\phi_1\right)\, ,\\
T^r{}_r =&\, T^\theta{}_\theta = T^\varphi{}_\varphi = \mu\left|\Dot{\phi_1}\right| - U\left( \phi_1 \right)\, .
\end{align}
Whereas the field equation \eqref{motion} for the scalar field reduces to:
\begin{align}
\frac{dU}{d\phi_1} = - 3\mu H\frac{\dot{\phi}_1}{\sqrt{\dot{\phi}^2_1}}\, . 
\label{eqmotioniso}
\end{align}
One should note that the term $\dot{\phi}_1/\sqrt{\dot{\phi}^2_1}$ in the r.h.s. of the equation provides a sign. 
Thus, with no presence of any other type of matter and a null cosmological constant, the FLRW equations are
\begin{align}
3 H^2 =&\, U\left( \phi_1 \right)\, , 
\label{frlweqiso1}\\ 
 -2 \dot{H}-3H^2 =&\, \mu \left|\Dot{\phi_1}\right| - U\left( \phi_1 \right)\, , 
\label{frlweqiso2}
\end{align}
and thus,
\begin{align}
 \left| \Dot{\phi_1}\right| = -\frac{2}{\mu}\Dot{H}= -\frac{2}{\mu} \left(\frac{\Ddot{a}(t)}{a(t)} - \frac{\Dot{a}^2(t)}{a(t)^2} \right)\, . 
\label{scalariso}
\end{align}
By \eqref{frlweqiso1}, the scalar field equation \eqref{eqmotioniso} can be expressed in the form:

\begin{align}
\left(\frac{dU}{d\phi_1}\right)^2 - 3\mu^2 U\left( \phi_1 \right) = 0\, . 
\label{edo}
\end{align}
Here, one of the main features of the cuscuton arises. 
Since the action for the cuscuton field is linear in the velocities, the scalar field equation 
\eqref{eqmotioniso} does not provide any information on the dynamics of the scalar field, and consequently, by an arbitrary potential 
$U\left( \phi_1 \right)$, the only possible solutions are provided by a constant scalar field $\phi_1=\phi_0$ that are the roots of the equation 
\eqref{edo}, playing the role of an effective cosmological constant. 
Nevertheless, that is not the case if one considers \eqref{edo} as a differential equation for the potential with respect to the cuscuton field. 
In such a case, by solving the equation (\ref{edo}), the following form for the potential is obtained
\begin{align}
U\left( \phi_1 \right) = \frac{3}{4}\mu^2 \left( \phi_1-\phi_0 \right)^2\, ,
\label{potentialcuscutonk=0}
\end{align}
where $\phi_{0}$ is an integration constant. 
It is straightforward to show that for this potential, the system of differential equations \eqref{frlweqiso1}-\eqref{frlweqiso2} is undetermined 
and the model does not provide any consistent and unique solution. 
However, as shown below, the cuscuton field in the presence of other types of matter might provide a way to regularize the cosmological expansion.\\

\begin{figure}[ht!]
 \centering
\includegraphics[scale=0.5]{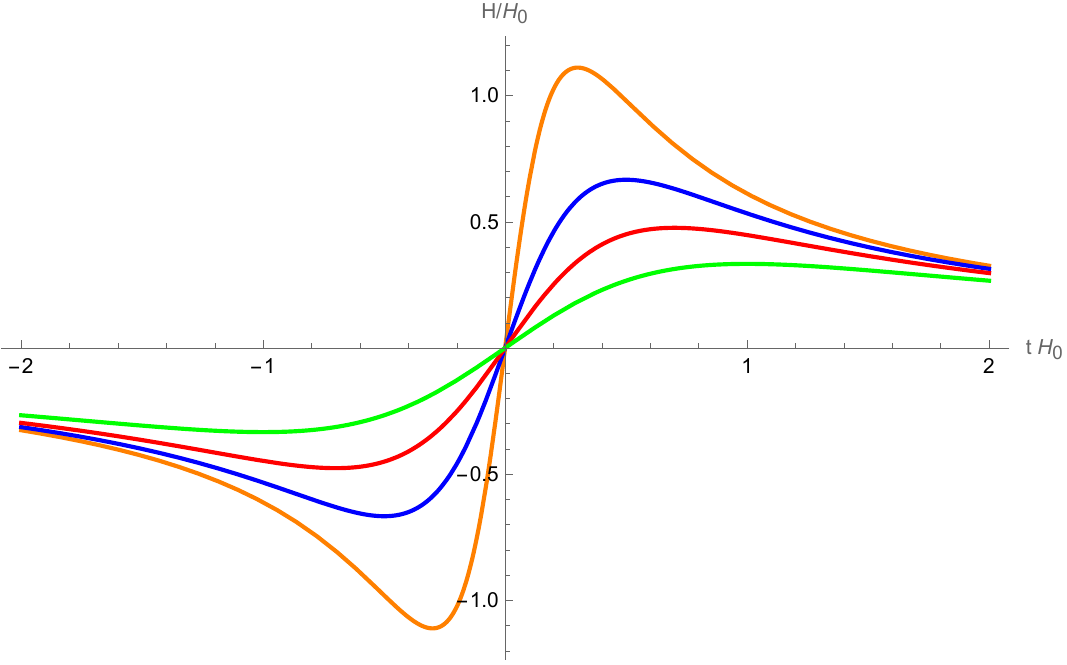} 
\caption{Hubble parameter for $b=0.3 H_0$ (orange), $b=0.5 H_0$ (blue), $b=0.7 H_0$ (red) and $b=1.0 H_0$ (green), and $\Omega_\mathrm{m}=0.3$.}
\label{Fig1b}
\end{figure}

\begin{figure}[ht!]
 \centering
\includegraphics[scale=0.4]{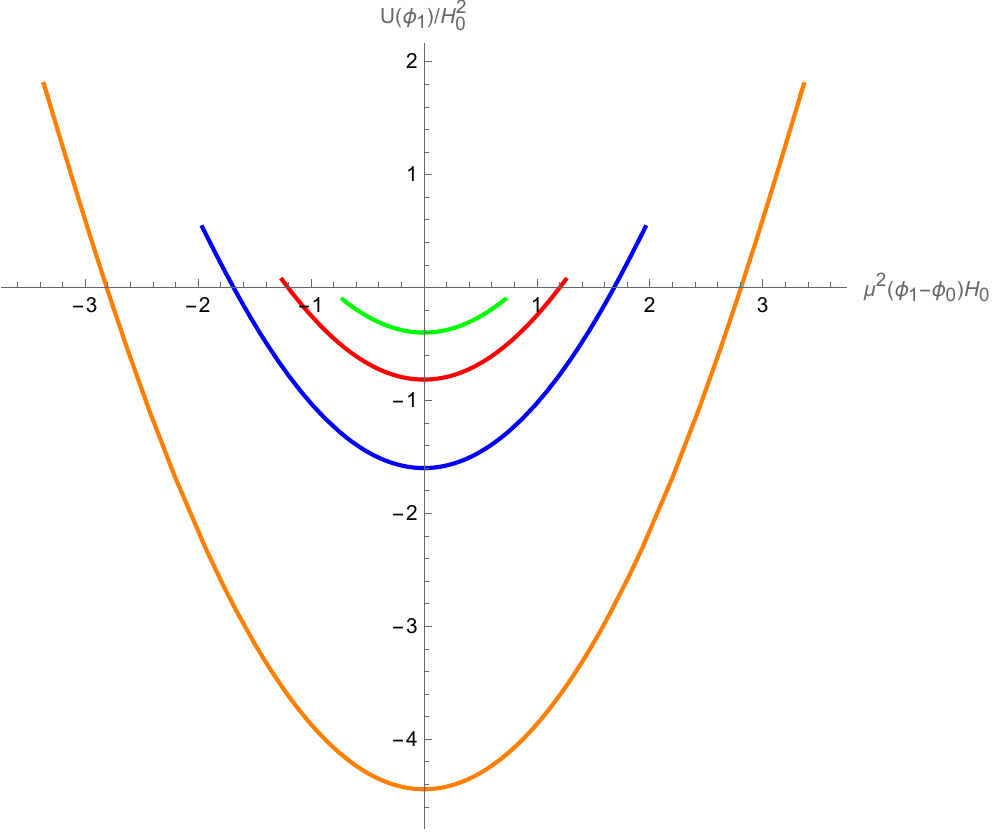} 
\caption{Potential of the cuscuton field $U\left( \phi_1 \right)$ for $b=0.3 H_0$ (orange), $b=0.5 H_0$ (blue), $b=0.7 H_0$ (red) and $b=1.0 H_0$ (green), and $\Omega_\mathrm{m}=0.3$.}
\label{Fig1}
\end{figure}

\begin{figure}[ht!]
 \centering
\includegraphics[scale=0.4]{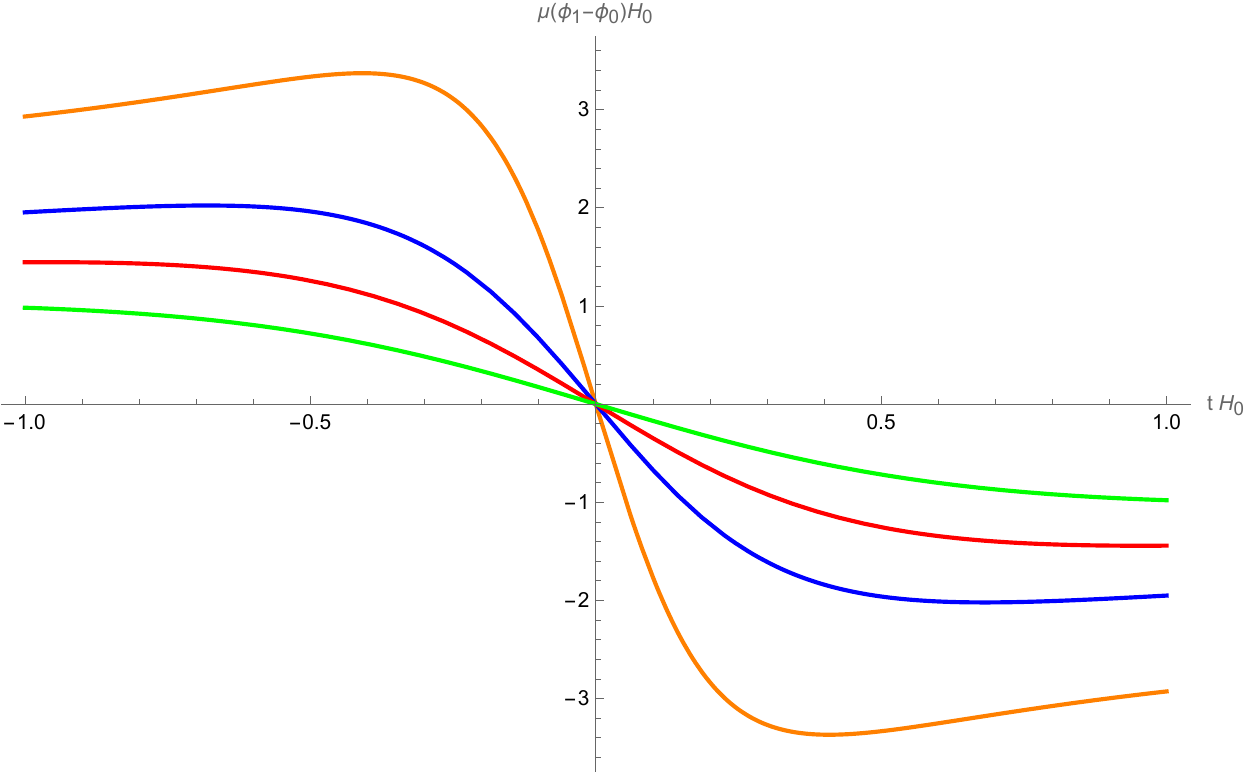} 
\caption{Evolution of the cuscuton field $\phi_1$ for $b=0.3 H_0$ (orange), $b=0.5 H_0$ (blue), $b=0.7 H_0$ (red) and $b=1.0 H_0$ (green), and $\Omega_\mathrm{m}=0.3$.}
\label{Fig1c}
\end{figure}

The FLRW metric in isotropic coordinates~\eqref{frlw} has a time-like singularity when $a(t) \rightarrow 0$. In the Big Bang model, this singularity occurs at $t=0$, marking the beginning of physical time. In order to obtain nonsingular cosmological solutions, the dependence of the scale factor on time is forced to undergo a Simpson-Visser-like regularisation:
\begin{align}
\Bar{t}^2 = t^2+b^2 \rightarrow \Bar{t} = \pm \sqrt{t^2+b^2} \, .
\end{align}
So the line element for this case is 
\begin{align}
ds^2 = -dt^2+a^2\left(\sqrt{t^2+b^2}\right) \left[dr^2+r^2 d\Omega^2 \right]\, ,
\end{align}
where $b\neq0$ is a constant, such that the scale factor reaches a minimum $a(t=0)=a_{\text{min}}\neq 0$ and the 
Big Bang singularity is avoided while the universe undergoes a bounce. 
Let us consider a FLRW spacetime with the presence of the cuscuton field and a perfect fluid with a constant equation of state, $p=w\rho$. Then, the FLRW equations turn out:
\begin{align}
3 H^2=&\, \kappa^2\rho + U\left( \phi_1 \right)\, ,\nn
 -2\Dot{H} =&\, \kappa^2\left(\rho+p\right) + \mu\left| \Dot{\phi_1} \right|\, .
\label{FLRWeqs3}
\end{align}
Then, the perfect fluid in terms of the scale factor evolves as 
\begin{align}
\rho=\rho_0\left(\frac{1}{a}\right)^{3(1+w)}\, .
\end{align}
For a given cosmological expansion $H(t)$, the corresponding potential and scalar field evolution is obtained by the FLRW equations \eqref{FLRWeqs3}
\begin{align}
U\left( \phi_1 \right) =&\, 3H^2 - \kappa^2 \rho_{0} a^{-3(1+w)}\, , \nn
\mu\left|\Dot{\phi_1}\right| =&\, -2\Dot{H} - \kappa^2\rho_{0} (1+w) a^{-3(1+w)}\, .
\label{phi}
\end{align}
One should note that according to the equation \eqref{phi}, the first derivative $\dot{\phi_1}$ must be a monotonically increasing or decreasing function 
in order to keep the equations consistent. Let us consider the case of dust matter $w_\mathrm{m}=0$. 
In the absence of the cuscuton field, the scale factor goes as $a\sim t^{2/3}$, which clearly contains a singularity at $t=0$. 
Nevertheless, this expansion rate might be regularized by considering the following bouncing solution:
\begin{align}
a(t)=&\, a_0\left(t^2+b^2\right)^\frac{1}{3}\, ,\nn
\rightarrow\quad H(t)=&\, \frac{2t}{3\left( t^2+b^2 \right)}\, .
\label{regularSol}
\end{align}
For $t\gg b$, the solution is basically the same as the one for a matter-dominated universe. However, for $t\sim 0$, the universe reaches a bounce and the Big Bang singularity is avoided, as shown in 
Fig.~\ref{Fig1b}, where the cosmological evolution experiences an accelerated phase around $t=0$ and then the matter-dominated universe is recovered. 
For this simple model, the equations \eqref{phi} can be solved and the scalar field and its potential lead to
\begin{align}
\mu\left(\phi_1(t)-\phi_{0}\right) =&\, -\frac{4}{3}\left[\frac{t}{t^2+ b^2}+\frac{\Omega_\mathrm{m}\arctan\left(\frac{t}{b}\right)}{b}\right]\, ,\nn
U\left( \phi_1 \right)=&\, \frac{4t^2-\Omega_\mathrm{m}\left( t^2+b^2 \right)}{3\left( t^2+b^2 \right)^2}\, ,
\label{potencialPhi1}
\end{align}
where $\Omega_\mathrm{m}=\frac{\rho_0}{\frac{3}{\kappa^2}H_0^2}$ and $\phi_{0}$ is a constant of integration. 
In Fig.~\ref{Fig1}, the potential for the cuscuton field is depicted. 
As shown, the smaller the values of the parameter $b$ are, the deeper the potential arises. 
This is natural, as $b$ is the central parameter that regularises the solution, and as goes to zero, the scalar field in \eqref{potencialPhi1} diverges and the singularity at $t=0$ is recovered. 
Moreover, in order to check the consistency of the second equation in \eqref{phi}, the behaviour of the scalar field is depicted in Fig.~\ref{Fig1c}. 
As shown, one has to consider $\mu<0$ in order to keep $\mu\left|\dot{\phi}_1\right|<0$ around the bounce, located at $t=0$, whereas away from the bounce $\mu\left|\dot{\phi}_1\right|>0$. 
Since this is not possible as $\mu$ is a constant, this might be interpreted as the cuscuton field playing a central role just around the bounce at $t=0$, where the solution is regularized. 
While away from $t=0$, the cuscuton field is not allowed and the matter-dominated universe is recovered. 
In addition, $\mu<0$ implies that the cuscuton field owns a negative kinetic energy and becomes a ghost, violating the energy conditions.
The elimination of the ghost is discussed in Section~\ref{ghostsection}.

In the section below, the same technique is applied to construct a non-stationary solution for the Ellis-Bronnikov wormhole.

%%%%%%%%%%%%%%%%%%%%%%%%%
\section{Non-stationary Ellis-Bronnikov wormhole}\label{NSEllisBronnikov}
%%%%%%%%%%%%%%%%%%%%%%%%

Let us start by reviewing the stationary Ellis-Bronnikov wormhole. Scuh a metric is a solution to Einstein's field equations in the presence of a massless scalar field 
with negative kinetic energy acting as the source. 
The spherically symmetric line element is typically expressed as
\begin{align}
ds^2 = -dt^2 + \frac{dr^2}{1-\dfrac{h(r)}{r}} + r^2d\Omega^2\, , 
\label{bronnikov}
\end{align}
where $h(r)$ is the shape function that governs the behavior of the wormhole. 
For the Ellis-Bronnikov wormhole, the shape function takes the particular form:
\begin{align}
h(r) = \frac{b^2}{r}\, ,
\end{align}
where $b$ is a constant. 
The metric \eqref{bronnikov} might be expressed more naturally through the coordinate transformation $r^2 = b^2 + \Bar{r}^2$, 
such that the line element (\ref{bronnikov}) becomes
\begin{align}
ds^2=-dt^2+d\Bar{r}^2+\Sigma^2d\Omega^2\, ,\quad \Sigma \left(\Bar{r}\right)=\sqrt{\Bar{r}^2+b^2}\, .
\label{metricEB2}
\end{align}
Here, we can see clearly that the metric is regular everywhere, and the radial coordinate $-\infty<\Bar{r}<\infty$ covers two universes/regions connected by a throat located at 
$\Bar{r}=0$, with a radius $R=b$. 
Thus, the non-vanishing components of the Einstein tensor are 
\begin{align}
G^{t}{}_t =&\, \frac{b^2}{\left(b^2+\Bar{r}^2\right)^2}\, , \nn
G^{r}{}_r =&\, -\frac{b^2}{\left(b^2+\Bar{r}^2\right)^2}\, , \nn
G^{\theta}{}_\theta =&\, G^{\varphi}{}_\varphi = \frac{b^2}{\left(b^2+\Bar{r}^2\right)^2}\, .
\end{align}
Let us now consider a scalar field described by the action:
\begin{align}
S_{\phi_2}=\int d^4x\sqrt{-g}\left[-\varepsilon g^{\mu\nu} \partial_\mu \phi_2\partial_\nu \phi_2-V\left( \phi_2 \right)\right]\, ,
\label{actionPhi1}
\end{align}
where $\varepsilon$ is a constant that can be either positive or negative. 
If $\varepsilon$ is negative, the scalar field $\phi_2$ 
becomes a ghost, and therefore the model turns out to be physically inconsistent. 
Since the metric is static and spherically symmetric, the scalar field by the field equations depends just on the radial coordinate, $\phi_2=\phi_2(r)$. 
Then, the expression of the energy-momentum tensor for the scalar field is given by
\begin{align}
T^\mu{}_\nu = 2\varepsilon g^{\mu \gamma}\partial_{\gamma}\phi_2\partial_{\nu}\phi_2 
 -\delta^{\mu}_{\nu}\left( \varepsilon g^{\alpha\beta}\partial_{\alpha}\phi_2\partial_{\beta}\phi_2-V(\phi) \right)\, , 
\label{tensormomentoenergiaescalar}
\end{align}
where its non-vanishing components are
\begin{align}
T^t{}_t =&\, -\varepsilon \phi_2'^2+V\left( \phi_2 \right)\, , \nonumber \\
T^r{}_r =&\, \varepsilon \phi_2'^2+V\left( \phi_2 \right)\, , \nonumber \\
T^\theta{}_\theta =&\, T^\varphi{}_\varphi = -\varepsilon \phi_2'^2+V\left( \phi_2 \right)\, . \nonumber 
\end{align}
Here, the primes denote differentiation with respect to the radial coordinate. 
By solving the field equations, one gets the solution: 
\begin{align}
V\left( \phi_2 \right)=0\, ,\quad \varepsilon=-1\, .
\end{align}
As $\varepsilon=-1$, the scalar field is a ghost and the model might not be physically unacceptable. 
While a first integral for the scalar field equation yields
\begin{align}
 \phi_2'= \frac{|b|}{\left(b^2+\Bar{r}^2\right)}\, , \label{scalarfield}
\end{align}
whose solution is given by 
\begin{align}
\label{phi2sol}
\phi_2=\phi_0 +\arctan \left(\frac{\Bar{r}}{b}\right)\, ,
\end{align}
where $\phi_{0}$ is an integration constant. Hence, the source for this wormhole is a free phantom scalar field. 
As far as $b\to 0$, the metric \eqref{metricEB2} turns out to be Minkowski spacetime and the scalar field becomes a constant that does not contribute to the equations, as expected. \\

Let us now reconstruct a non-stationary and regular wormhole, based on the Ellis-Bronnikov solution and the regularisation of FLRW spacetime followed above. 
To do so, we start with the Minkowski metric, which in spherical coordinates is given by 
\begin{align}
ds^2 = -dt^2 + dr^2 + r^2d\Omega^2 \, .
\label{flat}
\end{align}
Since the FLRW metric is conformal to Minkowski, it can be obtained just by multiplying the spatial part of Minkowski by the scale factor, 
\begin{align}
ds^2 = -dt^2 + a^2(t)\left(dr^2 + r^2d\Omega^2 \right)\, .
\label{frlw11}
\end{align}
On the other hand, by replacing the radius $r$ of the two-dimensional sphere $\mathcal{S}^2$ by $\Sigma = \sqrt{\Bar{r}^2+b^2}$ of \eqref{metricEB2} in \eqref{flat}, 
which is a kind of the Simpson-Visser regularisation procedure, one obtains 
\begin{align}
ds^2=-dt^2+ d\Bar{r}^2+\Sigma^2d\Omega^2 \, , 
\label{svprocedure}
\end{align}
which is the stationary Ellis-Bronnikov wormhole. 
The question now arises: how can we define a non-stationary wormhole? 
There are many possibilities. 
For example, if we apply the Simpson-Visser procedure directly on $\Bar{r}$ in Eq.~\eqref{frlw11}, the spacetime metric becomes 
\begin{align}
ds^2=-dt^2+a^2(t)\left(d\Bar{r}^2+\Sigma^2d\Omega^2\right)\, . 
\label{cosmologicalwormhole}
\end{align}
On the other hand, if the last term of Eq.~(\ref{svprocedure}) is multiplied by $a(t)$, then the spacetime turns out \eqref{cosmologicalwormhole} again. 
Hence, both procedures are equivalent, and this metric describes a non-stationary wormhole connecting two regions/universes 
$-\infty<\Bar{r}<\infty$, whose throat, with a topology of two-dimensional sphere $\mathcal{S}^2$, is not constant but evolves with time as $R(t)=a(t)b$. 
For this metric, the Einstein tensor is given by
\begin{align}
G^t{}_t =&\, \frac{b^2}{a(t)^2 \left(b^2 + \Bar{r}^2\right)^2}-\frac{3 \dot{a}^2(t)}{a(t)^2}\, , \label{E00I} \\
G^{\Bar{r}}{}_{\Bar{r}} =&\, -\frac{b^2}{a(t)^2 \left(b^2 + \Bar{r}^2\right)^2} - \frac{2 \ddot{a}(t)}{a(t)} - \frac{\dot{a}(t)^2}{a(t)^2}\, ,\label{E11I} \\
G^\theta{}_\theta =&\, G^\varphi{}_\varphi =\frac{b^2}{a(t)^2 \left(b^2 + \Bar{r}^2\right)^2} - \frac{2 \ddot{a}(t)}{a(t)} - \frac{\dot{a}^2(t)}{a(t)^2}\, .\label{E22I}
\end{align}
In order to reconstruct the corresponding matter Lagrangian that reproduces this solution, two scalar fields, as the ones analysed in previous sections, are considered. 
Then, the action is given by
\begin{align}
S = \int &\, d^4x \sqrt{-g} \left[ \varepsilon g^{\mu\nu} \partial_\mu \phi_2 \partial_\nu \phi_2 - V\left( \phi_2 \right) \right. \nn
&\, \left. + \mu \sqrt{-g^{\mu\nu}\partial_\mu \phi_1 \partial_\nu \phi_1} - U\left( \phi_1 \right) \right]\, .
\end{align}
That is, the above wormhole will be described exactly by the sum of these two scalar fields. 
The components of the energy-momentum tensor are shown below.

For the cuscuton field $\phi_1=\phi_1(t)$, the energy-momentum tensor is given by \eqref{tensormecuscuton}, whose components do not change, 
\begin{align}
T^{t}{}_t =&\, - U\left( \phi_1 \right)\, ,\nn
T^{\Bar{r}}{}_{\Bar{r}} =&\, T^{\theta}{}_\theta = T^{\varphi}{}_\varphi = \mu \left|\Dot{\phi_1}\right| - U\left( \phi_1 \right)\, .
\end{align}
Whereas the second scalar field $\phi_2=\phi_2\left(\Bar{r}\right)$ is described by the energy-momentum tensor \eqref{tensormomentoenergiaescalar} with 
$\varepsilon=-1$, that is, $\phi_2$ is a physically inconsistent ghost.
\begin{align}
T^{t}{}_t =&\, \frac{\phi'^2_{2}}{a(t)^2} + V\left( \phi_2 \right)\, ,\nn
T^{\Bar{r}}{}_{\Bar{r}} =&\, - \frac{\phi'^2_{2}}{a(t)^2} + V\left( \phi_2 \right)\, , \nn
T^\theta{}_\theta =&\, T^\varphi{}_\varphi = \frac{\phi'^2_{2}}{a(t)^2} + V\left( \phi_2 \right)\, .
\end{align}
The field equation for $\phi_2$ yields:
\begin{align}
\frac{{\phi_2}''}{a^2}+\frac{2\Bar{r}}{a^2 \left( {\Bar{r}}^2+b^2 \right)}{\phi_2}' +\frac{dV}{d\phi_2}=0\, .
\label{phi2eq}
\end{align}

\begin{figure*}[ht!]
\begin{subfigure}{.4\textwidth}
 \centering
\includegraphics[scale=0.4]{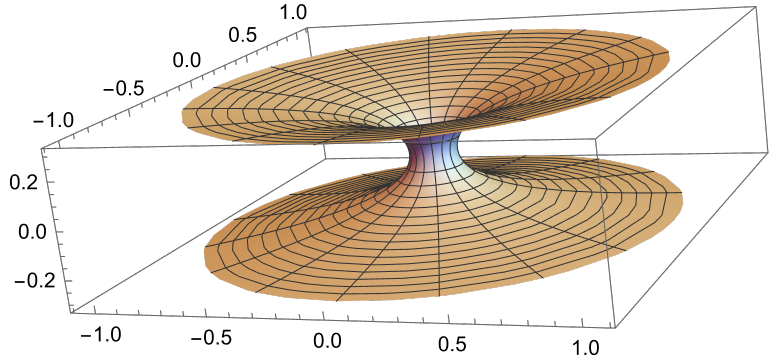}
 \caption{$t_0H_0=0$}
 \label{fig:sfig1}
\end{subfigure}%
%\begin{subfigure}{.4\textwidth}
% \centering
%\includegraphics[scale=0.4]{wormhole_t1}
% \caption{$t_0H_0=0.5$}
% \label{fig:sfig2}
%\end{subfigure}
%\begin{subfigure}{.4\textwidth}
% \centering
%\includegraphics[scale=0.4]{wormhole_t2}
% \caption{$t_0H_0=1$}
% \label{fig:sfig3}
%\end{subfigure}
\begin{subfigure}{.4\textwidth}
 \centering
\includegraphics[scale=0.4]{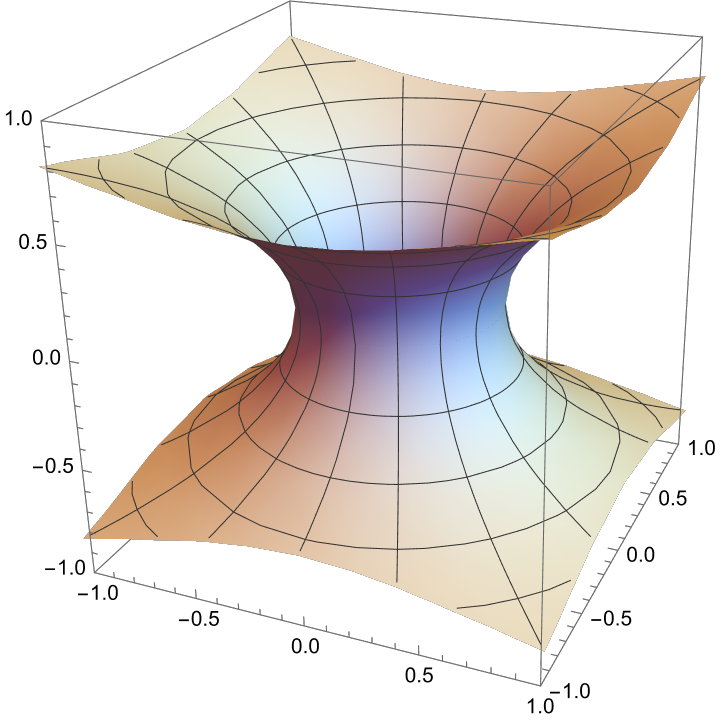}
 \caption{$t_0H_0=1.5$}
 \label{fig:sfig4}
\end{subfigure}
\caption{Evolution of the Ellis-Bronnikov wormhole for a de Sitter-like expansion $a(t)\propto \e^{tH_0}$ and $b=0.1$.}
\label{ExpoWormhole}
\end{figure*}

\begin{figure*}[ht!]
\begin{subfigure}{.4\textwidth}
 \centering
\includegraphics[scale=0.5]{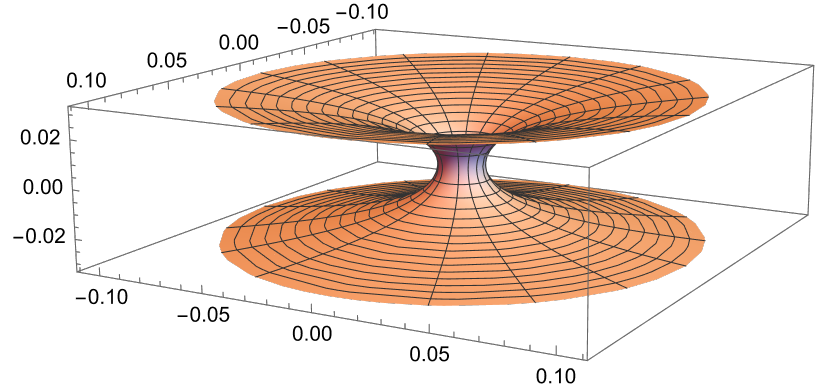}
 \caption{$\tilde{b}=0.1$}
 \label{fig:sfig1B}
\end{subfigure}%
\begin{subfigure}{.4\textwidth}
 \centering
\includegraphics[scale=0.6]{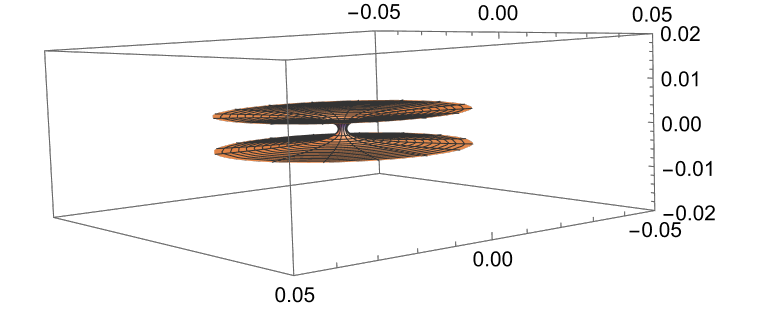}
 \caption{$\tilde{b}=0.01$}
 \label{ExpoWormhole1}
\end{subfigure}
\caption{Ellis-Bronnikov wormhole with $b=0.1$ for a dust-like scale factor $a(t)\propto (t^2+\tilde{b}^2)^{1/3}$ at $t=0$.}
\end{figure*}

For $V\left( \phi_2 \right)=0$, the scale factor $a(t)$ can be removed from \eqref{phi2eq} and the solution of the equation reproduces the same one as in the stationary 
Ellis-Bronnikov wormhole, i.e.
\begin{align}
\phi_2(r)=\phi_0+\arctan \left(\frac{\Bar{r}}{b}\right)\, .
\label{phi2SolNS}
\end{align}
Then, by the Einstein field equations, 
\begin{align}
G^t{}_t =&\, T^t{}_t\, , \quad G^{\Bar{r}}{}_{\Bar{r}} = T^{\Bar{r}}{}_{\Bar{r}}\, ,\nn
G^\theta{}_\theta =&\, T^\theta{}_\theta\, , \quad G^\varphi{}_\varphi = T^\varphi{}_\varphi\, ,
\label{EEcaseNS}
\end{align}
the general solution can be obtained, since the equations~\eqref{EEcaseNS} are separable in the radial and temporal parts, 
and the radial part fulfils the solution for $\phi_2$ given in \eqref{phi2SolNS}. 
Nevertheless, the temporal side of the equations, which implies the cuscuton field, reduces to the same equations 
as in the cosmological case analysed above, just with the presence of the cuscuton, which are given by: 
\begin{align}
3H^2=U\left( \phi_1 \right)\, , \quad -2\dot{H}-3H^2=\mu\left|\dot{\phi}_1\right| -U\left( \phi_1 \right)\, .
\label{timesideeqs}
\end{align}
As pointed out in Section~\ref{timeregu}, the only possible solution for an arbitrary potential is given by the roots $\phi_1=\phi_{0}=\mbox{constant}$ of the equation:
\begin{align}
\left(\frac{dU}{d\phi}\right)^2 - 3\mu^2 U(\phi) = 0\, . 
\label{rootsdS}
\end{align}
For such a case, the scale factor grows exponentially,
\begin{align}
a(t)\propto\e^{\sqrt{\frac{U(\phi_0)}{3}}t}\, .
\label{exponential}
\end{align}
For this case, the metric \eqref{cosmologicalwormhole} is free of timelike singularities and describes a wormhole whose throat grows exponentially with time, as shown in 
Fig.~\ref{ExpoWormhole}, where the structure of the wormhole is depicted at the equatorial plane, $\theta=\pi/2$. 
This is the only possible regular solution with no presence of any other kind of fields. 
Of course, as far as one includes additional species in the matter Lagrangian, as dust, for instance, the solution might be the one found in 
Section~\ref{timeregu} with the presence of $\rho_\mathrm{m}$, where the scale factor evolutes as $a(t)\propto(t^2+\tilde{b}^2)^{1/3}$. 
Note also that one might construct a non-stationary wormhole in the absence of the cuscuton field but with the presence of dust matter or radiation. 
However, in such cases, the spacetime metric would contain a time-like singularity, since the scale factor becomes null at $t=0$. 
This feature can be seen in Fig.~\ref{ExpoWormhole1}, as $\tilde{b}$ gets smaller, the throat of the wormhole is shrinking at $t=0$.
 
%%%%%%%%%%%
%%%%%%%%%%

%%%%%%%%%%%%%%%%%%%%%%%
\section{Eliminating ghosts}\label{ghostsection}
%%%%%%%%%%%%%%%%%%%%%%%

We have found that the cuscuton-type scalar field $\phi_1$ in (\ref{actionCusco}) is a ghost because $\mu$ is negative and 
another scalar field $\phi_2$ in (\ref{actionPhi1}) is also a ghost because $\varepsilon$ is negative. 
In \cite{Nojiri:2023dvf}, it has been shown that such ghosts can be eliminated by using the constraints. 
In this section, we discuss how the ghost is eliminated. 

We assume the solutions of the model as $\phi_1=\phi_1(t)=\Phi_1(t)$ as in (\ref{potencialPhi1}), 
$\phi_2=\phi_2\left( \Bar{r} \right) = \Phi_2 \left( \Bar{r} \right)$ as in (\ref{phi2sol}) or (\ref{phi2SolNS}). 
Here $\Phi_1(t)$ and $\Phi_2 \left( \Bar{r} \right)$ are not field variables but given functions of $t$ and $\Bar{r}$, respectively. 
If $\phi_1$ is a constant as in (\ref{rootsdS}), we can just drop the kinetic term of $\phi_1$ and therefore the cuscuton-type scalar field $\phi_1$ becomes non-dynamical and 
$\phi_1$ is not a ghost. 

For the FLRW universe, we add the following action to the action (\ref{actionCusco})\, 
\begin{align}
\label{cnstrnt1}
S_{\mathrm{cnstrnt}\, \phi_1} = \int d^4 x \sqrt{-g} \lambda_1 \left( g^{\mu\nu} \partial_\mu \hat\phi_1 \partial_\nu \hat\phi_1 + 1 \right)\, .
\end{align}
Here $\lambda_1$ is a Lagrange multiplier field and $\hat\phi_1$ is defined by $\phi_1=\Phi_1 \left( \hat\phi_1 \right)$. 
The variation of the action (\ref{cnstrnt1}) with rspect to $\lambda_1$, we obtain the following constraint for $\hat\phi_1$ and therefore $\phi_1$, 
\begin{align}
\label{cnstrnt1B}
0 = g^{\mu\nu} \partial_\mu \hat\phi_1 \partial_\nu \hat\phi_1 + 1 \, .
\end{align}
Due to the constraint \eqref{cnstrnt1B}, $\hat\phi_1$ and therefore $\phi_1$ become dynamical, that is, they do not propagate and therefore $\phi_1$ is not a ghost. 
The solution of the constraint (\ref{cnstrnt1B}) is given by $\hat\phi_1=t$, which gives $\phi_1=\Phi_1(t)$, consistently. 

For the Ellis-Bronnikov wormhole \eqref{metricEB2}, we add the following action to the action (\ref{actionPhi1}), 
\begin{align}
\label{cnstrnt2}
S_{\mathrm{cnstrnt}\, \phi_2} = \int d^4 x \sqrt{-g} \lambda_2 \left( g^{\mu\nu} \partial_\mu \hat\phi_2 \partial_\nu \hat\phi_2 - 1 \right)\, .
\end{align}
Here $\lambda_2$ is also a Lagrange multiplier field and $\hat\phi_2$ is defined by $\phi_2=\Phi_2 \left( \hat\phi_2 \right)$. 
The variation of the action (\ref{cnstrnt2}) with rspect to $\lambda_2$, we obtain the following constraint for $\hat\phi_2$ and therefore $\phi_2$, 
\begin{align}
\label{cnstrnt2B}
0 = g^{\mu\nu} \partial_\mu \hat\phi_2 \partial_\nu \hat\phi_2 - 1 \, .
\end{align}
Due to the constraint (\ref{cnstrnt2B}), $\hat\phi_2$ and therefore $\phi_2$ also become dynamical, that is, they do not propagate, and therefore $\phi_2$ is not a ghost. 
The solution of the constraint (\ref{cnstrnt2B}) is given by $\hat\phi_2=\Bar{r}$, which gives $\phi_2=\Phi_2 \left( \Bar{r} \right)$, consistently. 

For the time-dependent Simpson-Visser space time \eqref{cosmologicalwormhole}, we need to be technical a little bit, 
Because we are considering the solution that $\phi_1$ is a constant, we may eliminate the kinetic term of $\phi_1$ so that the cuscuton-type scalar field 
$\phi_1$ becomes non-dynamical and $\phi_1$ is not a ghost. 
Although $\phi_1$ is a constant, in order to include the field corresponding to the cosmological time $t$, we need to include the field $\hat\phi_1$ and the action \eqref{cnstrnt1}. 
Furthermore, the action \eqref{cnstrnt2} is modified to be 
\begin{align}
\label{cnstrnt3}
S_{\mathrm{cnstrnt}\, \phi_2}^{(2)} = \int d^4 x \sqrt{-g} \lambda_2 \left(\frac{1}{a\left(\hat\phi_1\right)} g^{\mu\nu} \partial_\mu \hat\phi_2 \partial_\nu \hat\phi_2 - 1 \right)\, , 
\end{align}
which gives the following constraint
\begin{align}
\label{cnstrnt2C}
0 = \frac{1}{a\left(\hat\phi_1\right)}g^{\mu\nu} \partial_\mu \hat\phi_2 \partial_\nu \hat\phi_2 - 1 \, .
\end{align}
Then by using $\phi_1=t$, which is a solution of \eqref{cnstrnt1B}, and the metric in \eqref{cosmologicalwormhole}, we find that $\hat\phi_2=\Bar{r}$ is a solution. 

We should note that we can consider the solution where $\lambda_1=\lambda_2=0$. 
Then the solutions in the previous sections are solutions even if we include the actions \eqref{cnstrnt1}, and \eqref{cnstrnt2} or \eqref{cnstrnt2B}.
Because both $\phi_1$ and $\phi_2$ become non-dynamical, the ghosts are excluded.

%%%%%%%%%%%%%%%%%%%%%
%%%%%%%%%%%%%%%%%%%%%

%%%%%%%%%%%%%%%%%%%%%%%
\section{The case in $f(R)$ gravity}\label{FRsection}
%%%%%%%%%%%%%%%%%%%%%%%

We might consider now the cases analysed above within $f(R)$ gravity, whose general action is given by
\begin{align}
S_G=\int d^4x \sqrt{-g} f(R)\, ,
\label{FRaction}
\end{align}
where $ f(R)$ is an arbitrary non-linear function of the Ricci scalar. 
The general form of the field equations is obtained by varying the action \eqref{FRaction} with respect to the metric, which, including the matter fields, leads to
\begin{align}
f_R R_{\mu\nu}-\frac{1}{2}g_{\mu\nu}f(R)-\left(\nabla_{\mu}\nabla_{\nu}-g_{\mu\nu}\Box\right)f_R=\kappa^2 T_{\mu\nu}^{\mathrm{matter}}\, .
\end{align}
Here $f_{R}= \frac{d f}{d R}$. 
This type of modified gravity is known to carry an extra scalar mode, such that the action might be rewritten as,% \cite{},
\begin{align}
S_G=\int d^4x \sqrt{-g} \left(\psi R-V(\psi)\right)\, .
\label{FRactionScalar}
\end{align}
Then, by varying this action and comparing \eqref{FRaction} and \eqref{FRactionScalar}, it yields
\begin{align}
\psi=f_{R}\, , \quad V(\psi)=f_{R} R-f(R)\, .
\label{phiR}
\end{align}
The field equations might be expressed now in an Einstein-like form as follows
\begin{align}
G_{\mu\nu}=\frac{1}{\psi}\left(T_{\mu\nu}^{(m)}+T_{\mu\nu}^{\psi}\right)\, ,
\label{fieldEqs}
\end{align}
where
\begin{align}
T_{\mu\nu}^{\psi}=\nabla_{\mu}\nabla_{\nu}\psi-g_{\mu\nu}\Box\psi-\frac{1}{2}g_{\mu\nu}V(\psi)\, .
\label{Tpsi}
\end{align}
We shall now find whether the above regular solutions, both cosmological as the Ellis-Bronnikov wormhole, can be reproduced in this type of theories. 
For the FLRW metric \eqref{frlw}, the only two independent equations are the $tt-$ and $rr-$ components in \eqref{fieldEqs}, which are given by
\begin{align}
3H^2+3H\frac{\dot{\psi}}{\psi}-\frac{1}{2}\frac{V(\psi)}{\psi}=&\, -\kappa^2T^{(m)t}{}_t\, , \nn
 -2\frac{\ddot{a}}{a}\psi-H^2-2H\dot{\psi}+\frac{1}{2}\frac{V(\psi)}{\psi}=&\, \kappa^2T^{(m)r}{}_r \, .
\label{eqsPhiFLRW}
\end{align}
For the sake and with no loss of generality, we can assume here $T_{\mu\nu}^{m}=0$, i.e., no presence of any other fields but just vacuum solutions. 
At the end, for a given cosmological solution $a(t)$, equations \eqref{eqsPhiFLRW} are differential equations with respect to the auxiliary field $\psi$ and its potential 
$V(\psi)$, whereas matter terms provide some particular solutions. 
We might consider the previous case of a scalar factor that mimics a dust-dominated universe, $a(t)\sim t^{2/3}$. 
Equations~\eqref{eqsPhiFLRW} can be solved, which yield \cite{Nojiri:2003ft,Saez-Gomez:2008ren},
\begin{align}
\psi(t) =&\, c_1 t^\frac{5+\sqrt{73}}{6}+c_2 t^\frac{5-\sqrt{73}}{6}\, ,\nn
V(t) =&\, c_{+}t^\frac{-7+\sqrt{73}}{6}+c_{-}t^\frac{-7-\sqrt{73}}{6}\, ,
\end{align}
where $c_{\pm}=\frac{2(9\pm\sqrt{73})}{3}c_{1,2}$. 
By inverting the expressions \eqref{phiR}, one finally gets the following $f(R)$ gravity:
\begin{align}
f(R)=\tilde{c}_1 R^{n_{+}}+\tilde{c}_2 R^{n_{-}}\, .
\end{align}
Here $n_{\pm}=\frac{7\pm\sqrt{73}}{12}$ and $\tilde{c}_{1,2}$ are just redefinitions of the integration constants. 
One might consider the same cosmological solution in the presence of dust matter and the cuscuton field, 
\begin{widetext}
\begin{align}
\psi = t^{\frac{5+\sqrt{73}}{6}}\left[c_1-\frac{3}{\sqrt{73}}\int \left(\kappa ^2 \rho_0 t^{-\frac{11+\sqrt{73}}{6}}+\mu ^2 t^{\frac{1-\sqrt{73}}{6}} \dot{\phi}\right)dt\right] %\nn
+ t^{\frac{5-\sqrt{73}}{6}}\left[c_2+\frac{3}{\sqrt{73}}\int \left(\kappa ^2 \rho_0 t^{\frac{-11+\sqrt{73}}{6}}+\mu ^2 t^{\frac{1+\sqrt{73}}{6}} \dot{\phi}\right)dt\right]\, .
\end{align}
\end{widetext}
Then, with the combination of the cuscuton equation \eqref{eqmotioniso} and the cosmological equations \eqref{eqsPhiFLRW}, the corresponding $f(R)$ action can be reconstructed. 
Nevertheless, one should note that the system of equations is undetermined, since there are three independent equations and four variables. 
In any case, the solution $a(t)\sim t^{2/3}$ can be well reproduced in $f(R)$ gravity, as shown in previous literature. 
We might try now to find out whether the regularized version of this cosmological solution, $a(t)\sim\left( t^2+b^2 \right)^{1/3}$, is a solution of $f(R)$ gravity, as it is in the case in 
GR with the presence of the cuscuton field. 
For such a case, the equations \eqref{eqsPhiFLRW} in vacuum provide the following solution for the auxiliary scalar field,
\begin{align}
\psi(t)=\left( t^2+b^2 \right)^{2/3}\left[c_1 \mathcal{P}^{k}_{j}(i t/b)+c_2\mathcal{Q}^{k}_{j}(i t/b)\right]\, .
\label{solRegularFR}
\end{align}
Here, $ \mathcal{P}^{k}_{j}$ and $\mathcal{Q}^{k}_{j}$ are the associated Legendre polynomials of the first and second kind, respectively, whereas 
$j=\frac{-3+\sqrt{73}}{6}$ and $k=\frac{2\sqrt{10}}{3}$. 
This is not a real solution, and consequently, the gravitational action $f(R)$ turns out complex, and it cannot be considered a consistent theory of gravity. 
Moreover, even by considering the presence of the cuscuton field and/or dust matter, the only real-valued gravitational action that can be obtained is the Hilbert-Einstein action. 
Indeed, the cuscuton and any other matter fields just provide particular solutions to \eqref{solRegularFR}, and the general solution becomes complex as well. 
In addition, the same result is found for any regular solution of the type $a(t)\sim \left( t^2+b^2 \right)^{n}$. 
Hence, one can conclude that this type of regular solution is exclusive of General Relativity (with the presence of the appropriate matter fields). 

For the stationary wormhole \eqref{metricEB2}, one reaches a similar conclusion in the cosmological case. 
The effective tensor \eqref{Tpsi} yields:
\begin{align}
T^{(\psi)t}{}_t =&\, -\frac{1}{\psi}\left(\psi''+\frac{2r\psi'}{r^2+b^2}+\frac{1}{2}V(\psi)\right)\, ,\nn
T^{(\psi)r}{}_r=&\, -\frac{1}{\psi}\left(\frac{2r\psi'}{r^2+b^2}+\frac{1}{2}V(\psi)\right)\, .\nn
T^{(\psi)\theta}{}_\theta=&\, T^{(\psi)\varphi}{}_\varphi=-\frac{1}{\psi}\left(\psi''+\frac{r\psi'}{r^2+b^2}+\frac{1}{2}V(\psi)\right)\, .
\label{efectWorm}
\end{align}
We might or might not consider the scalar field given by \eqref{tensormomentoenergiaescalar}, but the result turns out to be the same. 
By combining the $tt-$ and $\varphi\varphi-$ field equations, it gives
\begin{align}
\frac{r}{r^2+b^2}\frac{\psi'}{\psi}=0\, .
\label{EQcons}
\end{align}
The only possible solution for this equation is basically $\psi=$constant. 
Since the Ricci scalar is not constant, $R=-\frac{2b^2}{(r^2+b^2)^2}$ and $\psi=d f/dR$, the only gravitational action that admits the Ellis-Bronnikov wormhole is again the Hilbert-Einstein action. 
Nevertheless, one might consider additional sources of matter to reproduce the Ellis-Bronnikov wormhole in $f(R)$ gravity, but at the price of increasing the number of degrees of freedom. 

In addition, the case of non-stationary wormholes described by the metric \eqref{cosmologicalwormhole} does not provide any new insights, 
but in such a case, equations just become much messier and results turn out the same, i.e., the Hilbert-Einstein action is the only one that admits 
this type of regular solutions unless additional matter sources are considered.\\

\begin{figure}[ht!]
\centering
\includegraphics[scale=0.4]{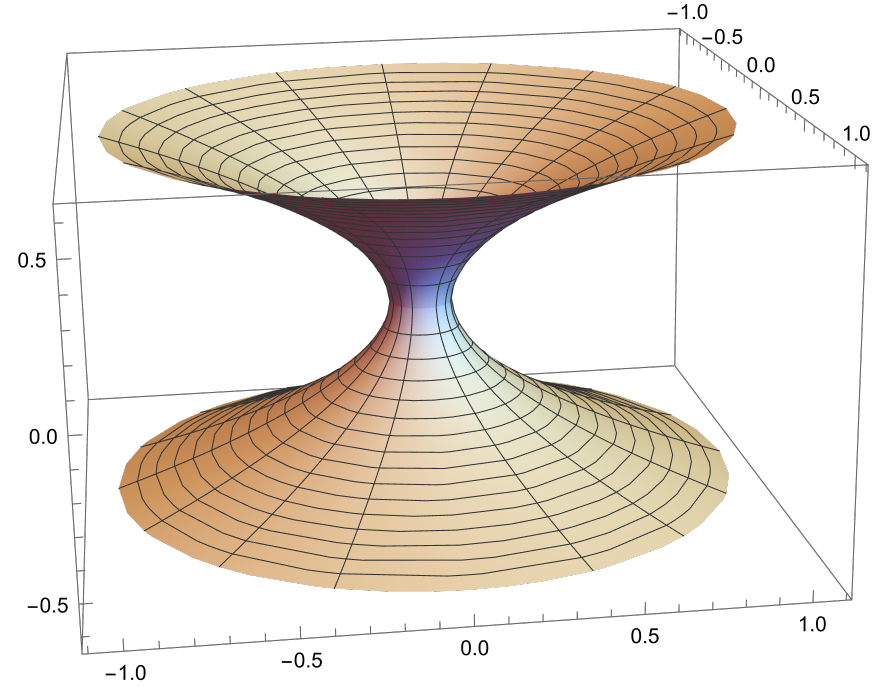} 
\caption{Wormhole within the framework of $f(R)$ gravity, described by the metric \eqref{lll2} for $r_0=0.1$.}
\label{FigFR}
\end{figure}

However, according to previous literature, some wormholes have been found in the framework of $f(R)$ gravity \cite{Godani:2018blx, Golchin:2019qch}. 
Then, we might explore the existence of similar spacetimes in comparison to the Ellis-Bronnikov wormhole to test in $f(R)$ gravity. 
To do so, let us start with the following generic spacetime metric, 
\begin{align}
\label{GBiv0}
ds^2 = - \e^{2\nu (r,t)} dt^2 + \e^{2\lambda (r,t)} dr^2 + r^2 d\Omega^2\, .
\end{align}
We might now consider the simplest wormhole geometry given by,
\begin{align}
\label{wh1} 
\e^{2\nu}=1\, , \quad \e^{2\lambda}=\frac{1}{1 - \frac{r_0}{r}}\, ,
\end{align}
where $r_0$ is the minimum radius of the wormhole and $r\geq r_0$. 
When $r \sim r_0$, we find $\e^{2\lambda}\sim \frac{r_0}{r - r_0}$, which looks singular at $r=r_0$. 
To avoid the singularity, if $r_0$ is a constant, by defining a new coordinate $l$, defined as follows,
\begin{align}
\label{lll}
r = \sqrt{ {r_0}^2 + l^2} \, ,\quad dr= \frac{ldl}{\sqrt{ {r_0}^2 + l^2}} \, ,
\end{align}
and then we obtain the following metric,
\begin{align}
\label{lll2}
ds^2=&\, - dt^2 + \frac{l^2}{{r_0}^2 + l^2 - r_0 \sqrt{ {r_0}^2 + l^2}} dl^2 \nonumber \\
&\, + \left( {r_0}^2 + l^2 \right) d\Omega^2\, . 
\end{align}
When $r\to r_0$, we find $l\to 0$, where $\frac{l^2}{{r_0}^2 + l^2 - r_0 \sqrt{ {r_0}^2 + l^2}} \to 2$, and therefore the spacetime is regular $r=r_0$, that is, $l=0$. 
We analytically continue the value of $l$ in the region where $l$ is negative. 
If the region with positive $l$ corresponds to our Universe, the region with negative $l$ corresponds to another Universe. 
The two Universes are connected by the wormhole, whose minimum radius is $r_0$. 
We should also note that when $\left|l\right|\to \infty$, we find $\left|l\right|\to r$ and $\frac{l^2}{{r_0}^2 + l^2 - r_0 \sqrt{ {r_0}^2 + l^2}} \to 1$ 
and therefore the spacetime is asymptotically flat. 
In Fig.~\ref{FigFR}, the structure of the wormhole is depicted at the equatorial plane, $\theta=\pi/2$, which shows a similar shape in comparison to the Ellis-Bronnikov wormhole. 

For the geometry in (\ref{wh1}), we find $R=0$, and the equations in \eqref{eqs1} have the following forms, 
\begin{align}
\label{eqs1B0}
& - \frac{1}{2} f(0) = - \frac{\kappa^2}{2} \rho \, , \quad 
\frac{1}{2} f(0) + \frac{r_0}{r^3} f'(0) = - \frac{\kappa^2}{2} p_\mathrm{rad} \, , \nn 
&\frac{1}{2} f(0) - \frac{r_0}{2r^3} f'(0) = - \frac{\kappa^2}{2} p_\mathrm{ang} \, , \quad 
0 = - \frac{\kappa^2}{2} j\, .
\end{align}
This tells that if $f(0)=f'(0)=0$, there appears the wormhole solution corresponding to \eqref{eqs1B0} even in the vacuum, where $\rho= p_\mathrm{rad} = p_\mathrm{ang} =0$. 
Because the square of the effective gravitational coupling is proportional to the inverse power of $f'$, the effective coupling diverges in this case where 
$f'(0)=0$ and therefore, the situation could not be realistic. 
We can consider, however, the expansion with respect to $f'(0)$ when $f'(0)$ does not vanish exactly, which could be regarded as a strong coupling expansion. 
Then, by the expansion, the geometry (\ref{eqs1B0}) can be deformed, but in the leading order, the geometry remains a solution. \\

This wormhole solution might also be extended to the non-stationary case by introducing a scale factor in front of the spatial part of the metric \eqref{lll2}. 
Nevertheless, in such a case, the Ricci scalar turns out to be $R=6\left(2H^2+\dot{H}\right)$, which depends solely on the time coordinate and consequently 
$f(R)$ so it does, whereas the field equations do not. 
Hence, one has to consider additional sources that will depend on both the time coordinate and the radial coordinate. 
This is an interesting open issue that might be explored in the future.

%%%%%%%%%%%%%%%%%%%%%%
\section{Conclusions}\label{conclusions}
%%%%%%%%%%%%%%%%%%%%%%%

In this work, we have explored the possibility of reconstructing non-stationary regular wormhole spacetimes by the use of two scalar fields, 
a usual phantom one that sources the wormhole and a cuscuton field that regularizes the time expansion of the wormhole. 
To do so, we have focused on a particular case, the so-called Ellis-Bronnikov wormhole, highlighting its theoretical significance and potential implications. 
Firstly, the Simpson-Visser procedure is applied to regularize, in a general way, the FLRW metric by introducing a bounce on the cosmological evolution that removes the 
Big Bang singularity at $t=0$, but does not cause significant changes in the sources. 
In particular, by applying such regularisation on a matter-dominated universe, the solution describes an eternal universe that is contracting 
till it reaches a minimum of the scale factor and then expands again. 
In such a process, an accelerating phase arises that might be identified with inflation \cite{Bartolo:2021wpt}, providing a regular cosmological model 
that might be well worth testing with the last observational data, which imposes severe constraints on the inflationary models. 
Such a regularization is possible due to the presence of the cuscuton field, whose main feature in comparison to usual scalar fields, is the linearity of its first derivative in the action. 
While cuscuton field just describes de Sitter-like expansion in vacuum, it makes cosmological solutions regular in the presence of ordinary matter. 

Nevertheless, the aim of the paper has lain in the reconstruction of regular non-stationary wormholes, particularly the 
Ellis Bronnikov wormhole, where the cosmological bouncing solution sourced by the cuscuton field is applied to do so. 
Then, a generalization of Ellis Bronnikov wormhole is proposed, where the spatial part of the metric evolves with time. 
The result is a non-stationary wormhole that, with the appropriate potential for the cuscuton field, it provides a completely regular spacetime, free of time-like and space-like singularities. 
Two particular cases are explored, one with an exponential expansion where the cuscuton field just acts as an effective cosmological constant and on the other hand, 
a dust matter-like expansion with the presence also of dust, where the cuscuton field plays an active role by regularizing the time evolution of the scale factor, 
which reaches a minimum value and then bounces. 
Several cases are depicted to show the time evolution and the features of this generalized non-stationary wormhole solution. 
Nevertheless, we have shown that when this type of regularisation is followed in the framework of 
GR by means of scalar fields, kinetic terms turn out negative and consequently, scalar fields become ghosts which might not be considered acceptable physically. 
Nevertheless, we have shown that ghost-free models can be reconstructed by using the constraints. 
 
In addition, the possibility of reproducing this type of non-stationary wormhole in a class of modified gravities, the so-called $f(R)$ gravity, is also explored. 
Results suggest that only within General Relativity with the presence of ghost scalar fields, the (non-)stationary 
Ellis-Bronnikov wormhole can be reconstructed, unless other types of anisotropic fluids for the matter sector are considered \cite{Lobo2009a}. 
Moreover, a novel wormhole solution in the framework of $f(R)$ gravity is obtained with similar features as the Ellis-Bronnikov solution but which does not require any ghost scalar field. 
This result suggests that other time-dependent black bounce metrics can be described in these theories without ghost fields. 
  
Further investigations might also point to a way for implementing a natural mechanism for the formation of such a kind of compact objects, one of the 
Achilles' heel when dealing with wormholes. Hence, future analysis may explore the implications of these scenarios in broader contexts. 

\section*{Acknowledgements}

This work is supported by the Spanish National Grant PID2020-117301GA-I00 funded by MICIU/AEI/10.13039/501100011033 (``ERDF A way of making Europe'' and ``PGC
Generaci\'{o}n de Conocimiento''); the Q-CAYLE project, funded by the European Union-Next Generation UE/MICIU/Plan de Recuperacion, 
Transformacion y Resiliencia/Junta de Castilla y Le\'{o}n (PRTRC17.11) and the financial support by the Department of Education, Junta de Castilla y Le\'{o}n and FEDER Funds, 
Ref.~CLU-2023-1-05.

\appendix

\section{}

For the spacetime, whose metric is given by, 
\begin{align}
\label{GBiv0B}
ds^2 =&\, - \e^{2\nu (r,t)} dt^2 + \e^{2\lambda (r,t)} dr^2 \nonumber \\
&\, + r^2 \left( d\theta^2 + \sin^2\theta d\phi^2 \right)\, ,
\end{align}
we find the following non-vanishing connections,
\begin{align}
\label{GBv}
&\Gamma^t_{tt}=\dot\nu \, , \quad \Gamma^r_{tt} = \e^{-2(\lambda - \nu)}\nu' \, , \quad \Gamma^t_{tr}=\Gamma^t_{rt}=\nu'\, , \nn
& \Gamma^t_{rr} = \e^{2\lambda - 2\nu}\dot\lambda \, ,\quad \Gamma^r_{tr} = \Gamma^r_{rt} = \dot\lambda \, , \nonumber \\
%&\nonumber \\
& \Gamma^r_{rr}=\lambda'\, , \Gamma^i_{jk} = \bar{\Gamma}^i_{jk}\, ,\quad \Gamma^r_{ij}=-\e^{-2\lambda}r \bar{g}_{ij} \, , \nn
&\Gamma^i_{rj}=\Gamma^i_{jr}=\frac{1}{r} \, \delta^i_{\ j}\, ,
\end{align}
where the metric $\bar{g}_{ij}$ is given by
$\sum_{i,j=1}^2 \bar{g}_{ij} dx^i dx^j = d\theta^2 + \sin^2\theta , d\phi^2$,
with $\left(x^1=\theta,\, x^2=\phi\right)$, and 
$\bar{ \Gamma}^i_{jk}$ represents the metric connection of $\bar{g}_{ij}$.
The ``dot'' and ``prime'' symbols denote differentiation with respect to $t$ and $r$, respectively.

Non-vanishing components of the curvatures are given by, 
\begin{widetext}
\begin{align}
\label{curvatures}
R_{rtrt} = & \, - \e^{2\lambda} \left\{ \ddot\lambda + \left( \dot\lambda - \dot\nu \right) \dot\lambda \right\}
+ \e^{2\nu}\left\{ \nu'' + \left(\nu' - \lambda'\right)\nu' \right\} \, ,\nonumber \\
R_{titj} =& \, r\nu' \e^{-2\lambda + 2\nu} \bar{g}_{ij} \, ,\nonumber \\
%&\nonumber \\
R_{rirj} =& \, \lambda' r \bar{ g}_{ij} \, ,\quad {R_{tirj}= \dot\lambda r \bar{ g}_{ij} } \, , \quad
R_{ijkl} = \left( 1 - \e^{-2\lambda}\right) r^2 \left(\bar{g}_{ik} \bar{g}_{jl} - \bar{g}_{il} \bar{g}_{jk} \right)\, ,\nonumber \\
%&\nonumber \\
R_{tt} =& \, - \left\{ \ddot\lambda + \left( \dot\lambda - \dot\nu \right) \dot\lambda \right\}
+ \e^{-2\lambda + 2\nu} \left\{ \nu'' + \left(\nu' - \lambda'\right)\nu' + \frac{2\nu'}{r}\right\} \, ,\nonumber \\
%&\nonumber \\
R_{rr} =& \, \e^{2\lambda - 2\nu} \left\{ \ddot\lambda + \left( \dot\lambda - \dot\nu \right) \dot\lambda \right\}
 - \left\{ \nu'' + \left(\nu' - \lambda'\right)\nu' \right\}
+ \frac{2 \lambda'}{r} \, ,\quad
R_{tr} =R_{rt} = \frac{2\dot\lambda}{r} \, , \nonumber \\
R_{ij} =&\, \left\{ 1 + \left\{ - 1 - r \left(\nu' - \lambda' \right)\right\} \e^{-2\lambda}\right\} \bar{g}_{ij}\, , \nonumber \\
%&\nonumber \\
R=& \, 2 \e^{-2 \nu} \left\{ \ddot\lambda + \left( \dot\lambda - \dot\nu \right) \dot\lambda \right\} 
+ \e^{-2\lambda}\left\{ - 2\nu'' - 2\left(\nu' - \lambda'\right)\nu' - \frac{4\left(\nu' - \lambda'\right)}{r} + \frac{2\e^{2\lambda} - 2}{r^2} \right\} \, .
\end{align}
\end{widetext}
By using the energy-momentum tensor $T_{\mu\nu}$, we now define the energy density $\rho$, the pressure in the radial direction $p_\mathrm{rad}$, 
the pressure in the angular direction $p_\mathrm{ang}$, and energy flow $j$ by, 
\begin{align}
\label{rhopj}
&T_{tt}=-\rho g_{tt}\, , \quad T_{rr} = p_{\parallel} g_{rr}\, , \quad 
T_{\theta\theta} = p_{\perp} g_{\theta\theta} \, , \nn 
&T_{\phi\phi}= p_{\perp} g_{\phi\phi} \, , \quad 
T_{tr}=T_{rt}= j \, .
\end{align}
The non-vanishing components of $f(R)$ equation, 
\begin{widetext}
\begin{align}
\label{JGRG13}
\frac{1}{2}g_{\mu\nu} f(R) - R_{\mu\nu} f'(R) - g_{\mu\nu} \Box f'(R) + \nabla_\mu \nabla_\nu f'(R)
= - \frac{\kappa^2}{2}T_{\mu\nu}\, ,
\end{align}
are given by
\begin{align}
\label{eqs1}
 - \frac{\kappa^2 \e^{2\nu}}{2} \rho=&\, - \frac{1}{2}\e^{2\nu} f(R) - \left[ - \left\{ \ddot\lambda + \left( \dot\lambda - \dot\nu \right) \dot\lambda \right\}
+ \e^{-2\lambda + 2\nu} \left\{ \nu'' + \left(\nu' - \lambda'\right)\nu' + \frac{2\nu'}{r}\right\} \right] f'(R) \nonumber \\
&\, + \e^{2\nu-2\lambda } \left\{ {\partial_r}^2 - \e^{2\lambda - 2\nu} \dot\lambda \partial_t - \lambda' \partial_r 
+ \frac{2}{r} \partial_r \right\} f'(R) \, , \nonumber \\
 - \frac{\kappa^2 \e^{2\lambda}}{2} p_{\parallel}&=\frac{1}{2}\e^{2\lambda} f(R) - \left[ \e^{2\lambda - 2\nu} \left\{ \ddot\lambda + \left( \dot\lambda - \dot\nu \right) \dot\lambda \right\}
 - \left\{ \nu'' + \left(\nu' - \lambda'\right)\nu' \right\} + \frac{2 \lambda'}{r} \right] f'(R) \nonumber \\
&\, - \left\{ - \e^{2\lambda -2\nu} \left( {\partial_t}^2 - \dot\nu \partial_t - \e^{-2\left(\lambda - \nu\right)}\nu' \partial_r \right)
+ \frac{2}{r} \partial_r \right\} f'(R) \, , \nonumber \\
&\, - \frac{\kappa^2 r^2}{2} p_{\perp}= \frac{r^2}{2} f(R) - \left\{ 1 + \left\{ - 1 - r \left(\nu' - \lambda' \right)\right\} \e^{-2\lambda}\right\} f'(R) \nonumber \\
&\, - r^2 \left\{ - \e^{-2\nu} \left( {\partial_t}^2 - \dot\nu \partial_t - \e^{-2\left(\lambda - \nu\right)}\nu' \partial_r \right)
+ \e^{-2\lambda} \left( {\partial_r}^2 - \e^{2\lambda - 2\nu} \dot\lambda \partial_t - \lambda' \partial_r\right) \right\} f'(R) \, , \nonumber \\
- \frac{\kappa^2}{2} j&= - \frac{2\dot\lambda}{r} f'(R) + \left( \partial_t \partial_r - \nu' \partial_t - \dot\lambda\partial_r \right) f'(R)\, .
\end{align}
\end{widetext}

%\bibliographystyle{unsrt}
%\bibliography{references} %%% Remove comment to use the external .bib file (using bibtex).


%\bibliographystyle{plain}

%\bibliography{012_Referencias}

\begin{thebibliography}{}

\bibitem{LIGOScientific:2016aoc}
B.~P.~Abbott, , \textit{et al.} [LIGO Scientific and Virgo],
%\textit{{Observation of Gravitational Waves from a Binary Black Hole Merger},
Phys. Rev. Lett., \textbf{116}(6), 061102 (2016);
%doi: \href{https://doi.org/10.1103/PhysRevLett.116.061102}{10.1103/PhysRevLett.116.061102}.
%\textit{{GW170817: Observation of Gravitational Waves from a Binary Neutron Star Inspiral},
Phys. Rev. Lett., \textbf{119}(16), 161101 (2017).
%doi: \href{https://doi.org/10.1103/PhysRevLett.119.161101}{10.1103/PhysRevLett.119.161101}.

\bibitem{EventHorizonTelescope:2022wkp}
K.~Akiyama \textit{et al.} [Event Horizon Telescope],
%\textit{{First Sagittarius A* Event Horizon Telescope Results. I. The Shadow of the Supermassive Black Hole in the Center of the Milky Way},
Astrophys. J. Lett., \textbf{930}(2), L12 (2022).
%doi: \href{https://doi.org/10.3847/2041-8213/ac6674}{10.3847/2041-8213/ac6674}.

\bibitem{Planck:2018vyg}
N.~Aghanim \textit{et al.} [Planck Collaboration],
%``Planck 2018 results. VI. Cosmological parameters,''
Astron. Astrophys. \textbf{641}, A6 (2020).

\bibitem{ACT:2020gnv}
S.Aiola \textit{et al.} [The Atacama Cosmology Telescope],
%\textit{{The Atacama Cosmology Telescope: DR4 Maps and Cosmological Parameters},
JCAP, \textbf{12}, 047 (2020).
%doi: \href{https://doi.org/10.1088/1475-7516/2020/12/047}{10.1088/1475-7516/2020/12/047}.

\bibitem{DeFalco2020}
V.~De Falco and E.~Battista and S.~Capozziello and M.~De Laurentis,
%\textit{General relativistic Poynting-Robertson effect to diagnose wormholes existence: static and spherically symmetric case},
Phys.~Rev.~D, \textbf{101}(10), 104037 (2020).
%doi: \href{https://doi.org/10.1103/PhysRevD.101.104037}{10.1103/PhysRevD.101.104037}.

\bibitem{Visser:1995cc}
M.~Visser, ``Lorentzian wormholes: From Einstein to Hawking,'' American Institute of Physics (1996)
%1077 citations counted in INSPIRE as of 16 Jun 2025
\bibitem{Lobo:2017book}
F.~S.~N.~Lobo,
%``Wormholes, Warp Drives and Energy Conditions,''
Fundam. Theor. Phys. \textbf{189} (2017), pp.-279
Springer (2017),
%ISBN 978-3-319-55181-4, 978-3-319-85588-2, 978-3-319-55182-1
%doi:10.1007/978-3-319-55182-1

\bibitem{morris1988wormholes}
M.~S.~Morris, and K.~S.~Thorne, and U.~Yurtsever,
%\textit{Wormholes, Time Machines, and the Weak Energy Condition},
Phys.~Rev. Letters, \textbf{61}(13), 1446--1449 (1988).
%doi: \href{https://doi.org/10.1103/PhysRevLett.61.1446}{10.1103/PhysRevLett.61.1446}.

\bibitem{RosenEinstein}
A.~Einstein and N.~Rosen,
%%\textit{{The Particle Problem in the General Theory of Relativity},
Phys. Rev., \textbf{48}, 73--77 (1935).
%doi: \href{https://doi.org/10.1103/PhysRev.48.73}{10.1103/PhysRev.48.73}.

\bibitem{wheeler1955geons}
J.~A.~Wheeler, 
%\textit{The Fate of the Einstein-Rosen Bridge},
Phys.~Rev., \textbf{97}(2), 511--536 (1955).
%doi: \href{https://doi.org/10.1103/PhysRev.97.511}{10.1103/PhysRev.97.511}.

\bibitem{instability_wormholes}
S.~Kojima,
%\textit{Blue-sheet instability and Schwarzschild wormholes},
Progress of Theoretical Physics, \textbf{73}, 1401-1416 (1985).
%doi: \href{https://doi.org/10.1143/PTP.73.1401}{10.1143/PTP.73.1401}.

\bibitem{Morris1988WormholesIS}
M.~S.~Morris and K.~S.~Thorne,
%\textit{Wormholes in spacetime and their use for interstellar travel: A tool for teaching general relativity},
American Journal of Physics, \textbf{56}, 395-412 (1988).
%Available at: \url{https://api.semanticscholar.org/CorpusID:119948364}.

\bibitem{visser2003traversable}
M.~Visser, and S.~Kar, and N.~Dadhich,
%\textit{Traversable wormholes with arbitrarily small energy condition violations},
Phys.~Rev. letters, \textbf{90}(20), 201102 (2003).

\bibitem{Lobo:2005us}
F.~S.~N.~Lobo,
%``Phantom energy traversable wormholes,''
Phys. Rev. D \textbf{71}, 084011 (2005).
%doi:10.1103/PhysRevD.71.084011
%[arXiv:gr-qc/0502099 [gr-qc]].

\bibitem{Lobo:2005yv}
F.~S.~N.~Lobo,
%``Stability of phantom wormholes,''
Phys. Rev. D \textbf{71}, 124022 (2005).
%doi:10.1103/PhysRevD.71.124022
%[arXiv:gr-qc/0506001 [gr-qc]].

\bibitem{Li:2020jyf}
A.~C.~Li and X.~F.~Li,,
%\textit{Morris-Thorne Wormhole in Vector-Tensor Theories with Abelian Gauge Symmetry Breaking},
Phys.~Rev.~D, \textbf{104}(4), 044006 (2021).
%doi: \href{https://doi.org/10.1103/PhysRevD.104.044006}{10.1103/PhysRevD.104.044006}.

\bibitem{Simpson_2019}
A.~Simpson, and M.~Visser,
%\textit{{Black-bounce to traversable wormhole},
JCAP, \textbf{02}, 042 (2019).
%doi: \href{https://doi.org/10.1088/1475-7516/2019/02/042}{10.1088/1475-7516/2019/02/042}.

\bibitem{lima2022black}
A.~M.~Lima, G.~M.~de Alencar Filho and J.~S.~Furtado Neto,
%``Black String Bounce to Traversable Wormhole,''
Symmetry \textbf{15}, no.1, 150 (2023)
%doi:10.3390/sym15010150
%[arXiv:2211.12349 [gr-qc]].

\bibitem{Furtado:2022tnb}
J.~Furtado, and G.~Alencar,
%\textit{{BTZ Black-Bounce to Traversable Wormhole},
Universe, \textbf{8}(12), 625 (2022).
%doi: \href{https://doi.org/10.3390/universe8120625}{10.3390/universe8120625}.

\bibitem{Lima:2023arg}
A.~Lima, G.~Alencar, R. N.~Costa Filho, and R.~R.~Landim, R. R.,
%\textit{{Charged black string bounce and its field source},
Gen. Rel. Grav., \textbf{55}(10), 108 (2023).
%doi: \href{https://doi.org/10.1007/s10714-023-03156-x}{10.1007/s10714-023-03156-x}.

\bibitem{Lima:2023jtl}
A.~Lima, and G.~Alencar, and D.~S\'aez-Chillon G\'omez,
%\textit{{Regularizing rotating black strings with a new black-bounce solution},
Phys. Rev. D, \textbf{109}(6), 064038 (2024).
%doi: \href{https://doi.org/10.1103/PhysRevD.109.064038}{10.1103/PhysRevD.109.064038}.

\bibitem{Alencar:2024yvh}
G.~Alencar, K.~A.~Bronnikov, M.~E.~Rodrigues, D.~S\'aez-Chill\'on G\'omez and M.~V.~de S.~Silva,
%``On black bounce space-times in non-linear electrodynamics,''
Eur. Phys. J. C \textbf{84}, no.7, 745 (2024).


\bibitem{Alencar:2025jvl}
G.~Alencar, A.~Duran-Cabac\'es, D.~Rubiera-Garcia and D.~S\'aez-Chill\'on G\'omez,
%``General spherically symmetric black bounces within nonlinear electrodynamics,''
Phys. Rev. D \textbf{111}, no.10, 104020 (2025).

\bibitem{Rois2025}
G.~I.~R{\'o}is, J.~T.~S.~S.~Junior, F.~S.~N.~Lobo and M.~E.~Rodrigues,
%``Novel electrically charged wormhole, black hole, and black bounce exact solutions in hybrid metric-Palatini gravity,''
Phys. Rev. D \textbf{111} (2025) no.12, 124012
%doi:10.1103/PhysRevD.111.124012
%[arXiv:2412.10324 [gr-qc]].

\bibitem{Radhakrishnan:2024rnm}
R.~Radhakrishnan, P.~Brown, J.~Mutulevich, E.~Davis, D.~Mirfendereski and G.~Cleaver,
%``A Review of Stable, Traversable Wormholes in f(R) Gravity Theories,''
Symmetry \textbf{16}, no.8, 1007 (2024)
%doi:10.3390/sym16081007
%[arXiv:2405.05476 [gr-qc]].

\bibitem{Elizalde:2023rds}
E.~Elizalde, S.~Nojiri, S.~D.~Odintsov and V.~K.~Oikonomou,
%``Propagation of gravitational waves in a dynamical wormhole background for two-scalar Einstein\textendash{}Gauss\textendash{}Bonnet theory,''
Phys. Dark Univ. \textbf{45}, 101536 (2024)
%doi:10.1016/j.dark.2024.101536
%[arXiv:2312.02889 [gr-qc]].

\bibitem{DeFalco2023}
V.~De Falco and S.~Capozziello,
%\textit{Static and spherically symmetric wormholes in metricaffine theories of gravity},
Phys.~Rev.~D, \textbf{108}(10), 104030 (2023).
%doi: \href{https://doi.org/10.1103/PhysRevD.108.104030}{10.1103/PhysRevD.108.104030}.




\bibitem{Garcia:2010xb}
N.~M.~Garcia and F.~S.~N.~Lobo,
%``Wormhole geometries supported by a nonminimal curvature-matter coupling,''
Phys. Rev. D \textbf{82}, 104018 (2010)
%doi:10.1103/PhysRevD.82.104018
%[arXiv:1007.3040 [gr-qc]].

\bibitem{Lobo:2008zu}
F.~S.~N.~Lobo,
%``General class of wormhole geometries in conformal Weyl gravity,''
Class. Quant. Grav. \textbf{25}, 175006 (2008)
%doi:10.1088/0264-9381/25/17/175006
%[arXiv:0801.4401 [gr-qc]].

\bibitem{Lobo2009a}
F.~S.~N.~Lobo and M.~A.~Oliveira,
%``Wormhole geometries in f(R) modified theories of gravity,''
Phys. Rev. D \textbf{80} (2009), 104012
%doi:10.1103/PhysRevD.80.104012
%[arXiv:0909.5539 [gr-qc]].

\bibitem{Harko2013}
T.~Harko, F.~S.~N.~Lobo, M.~K.~Mak and S.~V.~Sushkov,
%``Modified-gravity wormholes without exotic matter,''
Phys. Rev. D \textbf{87} (2013) no.6, 067504
%doi:10.1103/PhysRevD.87.067504
%[arXiv:1301.6878 [gr-qc]].

%
%\bibitem{wormholes_fr_gravity}
%R.~Lazkoz and Francisco~S.~N.~Lobo and others,
%%\textit{A Review of Stable, Traversable Wormholes in f(R) Gravity Theories},
%MDPI Universe, \textbf{7}, 1-27 (2021).
%%doi: \href{https://doi.org/10.3390/universe7080307}{10.3390/universe7080307}.

\bibitem{Ellis:1973yv}
H.~G.~Ellis,
%\textit{{Ether flow through a drainhole - a particle model in general relativity},
J. Math. Phys., \textbf{14}, 104--118 (1973).
%doi: \href{https://doi.org/10.1063/1.1666161}{10.1063/1.1666161}.

\bibitem{Bronnikov:1973fh}
K.~A.~Bronnikov,
%\textit{{Scalar-tensor theory and scalar charge},
Acta Phys. Polon. B, \textbf{4}, 251--266 (1973).

\bibitem{Nojiri:2023dvf}
S.~Nojiri and G.~G.~L.~Nashed,
%``Wormhole solution free of ghosts in Einstein{\textquoteright}s gravity with two scalar fields,''
Phys. Rev. D \textbf{108}, no.12, 124049 (2023)
%doi:10.1103/PhysRevD.108.124049
%[arXiv:2309.12379 [hep-th]].

\bibitem{Chataignier:2022yic}
L.~Chataignier, A.~Y.~Kamenshchik, A.~Tronconi and G.~Venturi,
%``Regular black holes, universes without singularities, and phantom-scalar field transitions,''
Phys. Rev. D \textbf{107}, no.2, 023508 (2023).

\bibitem{Bamba:2013fha}
K.~Bamba, A.~N.~Makarenko, A.~N.~Myagky, S.~Nojiri and S.~D.~Odintsov,
%``Bounce cosmology from $f(R)$ gravity and $f(R)$ bigravity,''
JCAP \textbf{01}, 008 (2014)
%doi:10.1088/1475-7516/2014/01/008
%[arXiv:1309.3748 [hep-th]].

\bibitem{Nojiri:2017ncd}
S.~Nojiri, S.~D.~Odintsov and V.~K.~Oikonomou,
%``Modified Gravity Theories on a Nutshell: Inflation, Bounce and Late-time Evolution,''
Phys. Rept. \textbf{692}, 1-104 (2017)
%doi:10.1016/j.physrep.2017.06.001
%[arXiv:1705.11098 [gr-qc]].

\bibitem{delaCruz-Dombriz:2018nvt}
\'A.~de la Cruz-Dombriz, G.~Farrugia, J.~L.~Said and D.~S\'aez-Chill\'on G\'omez,
%``Cosmological bouncing solutions in extended teleparallel gravity theories,''
Phys. Rev. D \textbf{97}, no.10, 104040 (2018)
%doi:10.1103/PhysRevD.97.104040
%[arXiv:1801.10085 [gr-qc]].

\bibitem{Afshordi:2006ad}
N.~Afshordi, D.~J.~H.~Chung and G.~Geshnizjani,
%``Cuscuton: A Causal Field Theory with an Infinite Speed of Sound,''
Phys. Rev. D \textbf{75}, 083513 (2007)
%doi:10.1103/PhysRevD.75.083513
%[arXiv:hep-th/0609150 [hep-th]].

\bibitem{cuscuton}
N.~Afshordi, D.~J.~H.~Chung, M.~Doran and G.~Geshnizjani,
%``Cuscuton Cosmology: Dark Energy meets Modified Gravity,''
Phys. Rev. D \textbf{75}, 123509 (2007)
%doi:10.1103/PhysRevD.75.123509
%[arXiv:astro-ph/0702002 [astro-ph]].


\bibitem{Bartolo:2021wpt}
N.~Bartolo, A.~Ganz and S.~Matarrese,
%``Cuscuton inflation,''
JCAP \textbf{05}, no.05, 008 (2022).

\bibitem{Dehghani:2025udv}
A.~Dehghani, G.~Geshnizjani and J.~Quintin,
%``Cuscuton bounce beyond the linear regime: bispectrum and strong coupling constraints,''
JCAP \textbf{05}, 026 (2025)
%doi:10.1088/1475-7516/2025/05/026
%[arXiv:2503.01992 [hep-th]].


\bibitem{Nojiri:2003ft}
S.~Nojiri and S.~D.~Odintsov,
%``Modified gravity with negative and positive powers of the curvature: Unification of the inflation and of the cosmic acceleration,''
Phys. Rev. D \textbf{68}, 123512 (2003)
%doi:10.1103/PhysRevD.68.123512
%[arXiv:hep-th/0307288 [hep-th]].

\bibitem{Saez-Gomez:2008ren}
D.~Saez-Gomez,
%``Modified f(R) gravity from scalar-tensor theory and inhomogeneous EoS dark energy,''
Gen. Rel. Grav. \textbf{41}, 1527-1538 (2009)
%doi:10.1007/s10714-008-0724-3
%[arXiv:0809.1311 [hep-th]].

\bibitem{Godani:2018blx}
N.~Godani and G.~C.~Samanta,
%``Traversable Wormholes and Energy Conditions with Two Different Shape Functions in $f(R)$ Gravity,''
Int. J. Mod. Phys. D \textbf{28}, no.02, 1950039 (2018)
%doi:10.1142/S0218271819500391
%[arXiv:1809.00341 [gr-qc]].

\bibitem{Golchin:2019qch}
H.~Golchin and M.~R.~Mehdizadeh,
%``Quasi-cosmological Traversable Wormholes in $f(R)$ Gravity,''
Eur. Phys. J. C \textbf{79}, no.9, 777 (2019)
%doi:10.1140/epjc/s10052-019-7292-4
%[arXiv:1908.04378 [gr-qc]]. 

\end{thebibliography}

\end{document}